\documentclass[%
 reprint,
 amsmath,amssymb,
 aps,
]{revtex4-2}

\usepackage{graphicx}% Include figure files
\usepackage{dcolumn}% Align table columns on decimal point
\usepackage{tikz}
\usepackage{bm}% bold math
\usepackage{amsmath}
\begin{document}

\preprint{APS/123-QED}

\title{A Geometric Approach to the Navier-Stokes Equations}% Force line breaks with \\

\author{Sebastián Alí Sacasa Céspedes$^{1}$}
 \altaffiliation[Also at ]{sebastian.sacasa@ucr.ac.cr; sebas201364@gmail.com}%Lines break automatically or can be forced with \\

\affiliation{%
$^1$Universidad de Costa Rica (UCR), San Pedro de Montes de Oca, San José, 11501-2060, Costa Rica}%

%\collaboration{}%\noaffiliation

\date{November 14, 2024}% It is always \today, today,

\begin{abstract}
\textbf{Introduction:} the Navier-Stokes equations are foundational in fluid dynamics, describing the motion of fluids such as liquids and gases. Despite their broad applications, analytically solving these equations—particularly for complex flows and high-Reynolds-number regimes—remains a significant challenge. While numerical simulations and partial solutions provide some insights, they often rely on restrictive assumptions, limiting their applicability. Recently, alternative geometric and algebraic methods have emerged, aiming to leverage the equations' underlying structure. However, critical questions regarding the uniqueness and stability of weak solutions remain unresolved, prompting conjectures and further research in this area. \textbf{Objective:} this paper aims to analyze the Navier-Stokes equations by reformulating them in covariant form and developing new equations that facilitate the search for potential solutions with a focus on symmetries. \textbf{Geometric Approach:} a covariant formulation of the Navier-Stokes equations was established to apply a Fourier transform on the bounded manifold while seeking smoothness and viable solutions via convergence of the element of manifold. \textbf{Transformations:} the equations and their transformations between manifolds were examined, investigating symmetries and potential interpretations, particularly with respect to homeomorphisms, isometries, and diffeomorphisms, including inertial frames of reference. \textbf{Discussion and Conclusions:} this study introduces a geometric reformulation of the Navier-Stokes equations, proposing new equations designed to enhance the convergence and smoothness of solutions. It presents a novel class of solutions, transformations, symmetries and interpretations, offering significant interdisciplinary connections. To broaden the applicability and scope of these equations and their solutions, further simulations, experimental validation, and continued development are essential.

\begin{description}
\item[Keywords] Navier-Stokes Equations, Manifold, Topological Metric Space, Macrotensors, Symmetries, Smoothness.

\end{description}
\end{abstract}

\maketitle

%\tableofcontents

\section{\label{sec:level1} Introduction}

The Navier-Stokes equations, a set of nonlinear partial differential equations, have been foundational to fluid dynamics. These equations describe the motion of fluid substances such as liquids and gases and have been extensively applied to model various natural phenomena, from ocean currents to atmospheric circulation and plasma physics. However, obtaining analytical solutions to the Navier-Stokes Equations, especially for complex flows and high-Reynolds-number regimes, remains an ongoing challenge.

Despite significant efforts, finding a general, closed-form solution to the Navier-Stokes Equations is still an open problem in both mathematics and physics. While numerical simulations and approximate methods have provided insights into certain scenarios, these approaches often rely on simplifying assumptions and are limited in scope. Partial solutions, including those derived from similarity solutions and asymptotic expansions, have illuminated specific aspects of fluid dynamics but remain constrained in their applicability.

In recent years, alternative approaches to solving the Navier-Stokes equations have emerged, including geometric and algebraic methods. These approaches aim to exploit the inherent structure of the equations, drawing from fields such as differential geometry, algebraic geometry, and theoretical physics. Weak solutions, which generalize the classical notion of solutions, have been developed to account for the irregularities and lack of smoothness typical in many fluid flows. However, questions of uniqueness and stability for these weak solutions continue to be the subject of ongoing research proposing conjectures such as Qualitative Regularity Conjecture, Local-in-Time Quantitative Regularity Conjecture and Global-in-Time Quantitative Regularity Conjecture [5].

This paper aims to analyze the Navier-Stokes equations by reformulating them in covariant form and developing new equations that facilitate the search for potential solutions with a focus on symmetries.  By modifying the geometric structure, it is intend to explore conditions that ensure convergence and smoothness of these solutions.

\section{\label{sec:level2}Geometric Approach}

According to the Navier-Stokes Equations may be written as follows using the Einstein Convention Summation [7].

\begin{equation}
\frac{\partial u_i}{\partial t} + u_j  \frac{\partial u_i}{\partial x_j}= \nu \frac{\partial^2 u_i}{\partial x_i^2} - \frac{\partial p}{\partial x_i} + f_i 
\end{equation}

\begin{equation}
    \frac{\partial u_i}{\partial x_i} = 0
\end{equation}

These equations are accomplished for a flat Euclidean geometry. 

Let suppose does exist a smooth manifold of dimension $D$ with boundaries and a metric microtensor o just a metric tensor $(M, g)$ according to [1]. The boundaries of the smooth manifold must exist in reason of the problem conditions. Using [4] equations (1) and (2) may be written as follows concerning just the components and not explicitly the base vectors, as they are implicit in the metric tensor.

\begin{equation}
\frac{\partial u_i}{\partial t} + u_j  \frac{\partial u_i}{\partial x_j}= \frac{\nu}{(\pm g)^{1/2}}\frac{\partial }{\partial x_k}[(\pm g)^{1/2} g^{ki} \frac{\partial u_i}{\partial x_i}] - \frac{\partial p_j}{\partial x_i} + f_i 
\end{equation}

\begin{equation}
    \frac{1}{(\pm g)^{1/2}} \ \frac{\partial }{\partial x_k} [(\pm g)^{1/2} u_k] = 0
\end{equation}

Developing the terms in both equations.

\begin{multline}
\frac{\partial u_i}{\partial t} + u_j  \frac{\partial u_i}{\partial x_j} \\= 
\frac{\nu}{(\pm g)^{1/2}} \Bigg[ (\pm g)^{1/2} \frac{\partial g^{ki}}{\partial x_k} \frac{\partial u_i}{\partial x_i} 
+ g^{ki} \frac{\partial^2 u_i}{\partial x_k \partial x_i}  \frac{1}{2} (\pm g)^{-1/2} \frac{\partial g}{\partial x_k} 
\\+ (\pm g)^{1/2} g^{ki} \frac{\partial u_i}{\partial x_i} \frac{1}{2} (\pm g)^{-1/2} \frac{\partial g}{\partial x_k} \Bigg] - \frac{\partial p_j}{\partial x_i} + f_i 
\end{multline}

\begin{multline}
\frac{\partial }{\partial x_k} [(\pm g)^{1/2} u_k] = \left[\frac{\partial }{\partial x_k}(\pm g)^{1/2}\right]  u_k + (\pm g)^{1/2}\frac{\partial u_k}{\partial x_k} \\ 
= \pm \frac{1}{2} (\pm g)^{-1/2} \frac{\partial g}{\partial x_k}  u_k + (\pm g)^{1/2}\frac{\partial  u_k }{\partial x_k} \\ 
= \gamma^{1/2} \left[\frac{1}{2} g^{-1/2} \frac{\partial g}{\partial x_k}  u_k + g^{1/2}\frac{\partial  u_k}{\partial x_k}\right] = 0
\end{multline}

Upon factoring and simplifying the expression, it is find that the relationship indicates that the rate of change is balanced by contributions from the metric tensor. This implies a sort of equilibrium where the influences from the motion via \( u_k \) and the space-time variations of the metric field captured by \( g \) are effectively counteracting each other. 

It is define a term \( A_{ij}(t) \) that collects the contributions from the second-order derivatives of the metric \( g \) and how \( u_i \) interacts with it. The formal definition involves viscosity \( \nu \) and components related to the gradients of \( g \).
\begin{multline}
A_{ij}(t) = A_{ij}[g^{ij}(x_{i}(t),x_{j}(t))] = \frac{\nu}{2\gamma g } \Bigg[ \left(\frac{\partial g^{ij}}{\partial x_j}\right)\left(\frac{\partial g}{\partial x_i}\right) 
\\- \frac{1}{2} g^{ji} \left(\frac{\partial g}{\partial x_j}\right)\left(\frac{\partial g}{\partial x_i}\right) + g^{ji}\left(\frac{\partial^2 g}{\partial x^2_i}\right) \Bigg]
\end{multline}
Finally, after factorize and replace the identical terms using (6) in (5), it is arrived at the simplified dynamic equation.

\begin{equation}
\frac{\partial u_i}{\partial t} + \gamma A_{ij}(t) u_i + \frac{\partial p_j}{\partial x_j} \frac{\partial x_j}{\partial x_i} - f_i = 0
\end{equation}

Being $\gamma = \pm 1$. The relation $\frac{\partial x_j}{\partial x_i} = \delta_j^i$ is the Kronecker Delta, changing the indexes for the pressure as follows. 

\begin{equation}
\frac{\partial u_i}{\partial t} + \gamma A u_i + \frac{\partial p_i}{\partial x_i} - f_i = 0 
\end{equation}

At this stage, it is possible to apply a Fourier Transform on the manifold and set the integrand to zero. This leads to the following equation, where it is leverage the Convolution Theorem to transition from position space to momentum space. In this context, the functions are treated as distribution operators that belong to a class $C^{k}$ of differentiable functions, utilizing the properties of the Dirac Delta function, [16, 43, 46, 48]. Yielding to the next equation.

\begin{multline}
[-ik_t + \gamma(2\pi)^{d+1}{A_{ij}(t)}] \hat{u_t} + i_{k_i} \hat{p_i} - \hat{f_i} = 0 \\
\Longrightarrow [-ik_t + \gamma(2\pi)^{d+1} A_{ij}(t)] \hat{u_t} = -i_{k_i} \hat{p_i} + \hat{f_i} \\
\Longrightarrow \hat{u_t} = [-i k_t + \gamma(2\pi)^{d+1} A_{ij}(t)]^{-1} [\hat{f_i}-i {k_i} \hat{p_i} ]
\end{multline}

Introducing a factor $\pm \epsilon i$ to move the pole at $k_{t}=0$ when $A_{ij}(t)$, which occurs when the derivatives of $g$ are set to zero in a manifold with zero curvature or when they do not have dependence of coordinates.

\begin{equation}
    \hat{u_t} = [\gamma((2\pi)^{d+1}A+\epsilon i)-ik_t]^{-1}[\hat{f}-ik_i \hat{p}]
\end{equation}

Setting the Inverse Fourier Transform.

\begin{multline}
  \int_{M}[\gamma((2\pi)^{d+1}A+\epsilon i)-ik_t]^{-1}\\ [\hat{f}-ik_i \hat{p}]e^{ikx}\frac{dk}{(2\pi)^{\frac{d+1}{2} }} 
\end{multline}
By integral properties this equation can be arrange as follows.
\begin{multline}
 I _M = [I_1]_M+ [I_2]_M \\= \int_{M} \frac{\hat{f}e^{ikx}}{[\gamma(2\pi)^{d+1}A_{ij}(t)+i(\gamma\epsilon -k_t)]}\frac{dk}{(2\pi)^{\frac{d+1}{2} }}\\-\int_{M}\frac{ik_i\hat{p_i}e^{ikx}}{[\gamma(2\pi)^{d+1}A+i(\gamma\epsilon -k_t)]} \frac{dk}{(2\pi)^{\frac{d+1}{2} }} 
\end{multline}
Making the substitution $Z=Z[A[g(x_i(t),x_j(t))] =\gamma(2\pi)^{d+1}A \not = Z(k)$ that implies that $I_M$ results in 
\begin{multline}
 I _M = [I_1]_M+ [I_2]_M = \int_{M}\frac{\hat{f}e^{ikx}}{[Z+i(\gamma\epsilon -k_t)]} \frac{dk}{(2\pi)^{\frac{d+1}{2} }} 
 \\- \int_{M} \frac{ik_i \hat{p}e^{ikx}}{[Z+i(\gamma\epsilon -k_t)]}\frac{dk}{(2\pi)^{\frac{d+1}{2} }}
\end{multline}
Both $[I_1]_M$ and $[I_2]_M$ exhibit discontinuities for $k_t$ as $\epsilon \rightarrow 0$, specifically at $k_t = Z$ and $k_t = 0$, as well as when $Z \rightarrow 0$. These integrals are of third kind. Nonetheless, using Feynman's prescription [6] and Cauchy's Integral Formula [56] to carry out the integration over the component of time of the frequency is possible to calculate the remainder.
\begin{multline}
  R_0=\lim_{k_t\rightarrow \gamma \epsilon -iZ}\frac{ k_t-( \gamma \epsilon -iZ)e^{-ik_t t}}{Z+i(\gamma\epsilon -k_t)} =  e^{-it(\gamma \epsilon -iZ)}  \\ \lim_{k_t\rightarrow \gamma \epsilon -iZ} \frac{ k_t-( \gamma \epsilon -iZ)}{Z+i(\gamma\epsilon -k_t)} \stackrel{L'H} = e^{-it(\gamma \epsilon -iZ)}  \lim_{k_t\rightarrow \gamma \epsilon -iZ} \frac{1}{-i}\\=   i e^{-it(\gamma \epsilon -iZ)} = i e^{Zt} = i e^{\gamma(2\pi)^{d+1}A_{ij}(t) t }   
\end{multline}
So the integral converges as
\begin{multline}
\int_M \frac{e^{-ik_t t}}{Z+i(\gamma\epsilon -k_t)} dk_t = 2\pi i R_0 = -2\pi e^{Zt} \\ =-2\pi e^{\gamma(2\pi)^{d+1}A_{ij}(t)t} =-2\pi \phi_{ij}(t) 
\end{multline}
Calling $\phi_{ij}(t)= \exp{[\gamma(2\pi)^{d+1}A_{ij}(t)t}]$. Replacing this result in the integration over the space coordinates developing the terms and making usage of Fubini's Theorem [10].
\begin{multline}
 u_j[x_j(t)]=-2\pi \phi_{ij}(t)  \int_{M} [\hat{f}-ik \hat{p}]e^{ikx}\frac{dk}{(2\pi)^{\frac{d+1}{2} }}\\ =  -2\pi \phi_{ij}(t)  [-\int_{M} ik \hat{p}\ e^{ikx}\frac{dk}{(2\pi)^{\frac{d+1}{2} }}+\int_{M}\hat{f}e^{ikx}\frac{dk}{(2\pi)^{\frac{d+1}{2} }}] 
 \\= -2\pi \phi_{ij}(t) [-\int_{M} ik (\int_Mp(x')e^{-ikx'}\frac{dx}{(2\pi)^{\frac{d+1}{2} }})\ e^{ikx}\frac{dk}{(2\pi)^{\frac{d+1}{2} }}\\+\int_{M}\hat{f}e^{ikx}\frac{dk}{(2\pi)^{\frac{d+1}{2} }}]
 \\ =-2\pi \phi_{ij}(t) [-\int_{M} \int_M ik p(x') \ e^{ik(x-x')}\frac{dk \ dx'} {(2\pi)^{d+1}}+f_i]\\=-2\pi \phi_{ij}(t) [-\int_{M} p(x') \frac{\partial \delta(x-x')}{\partial x} \frac{dx'} {(2\pi)^{\frac{d+1}{2}}}+f_i]
 \\= -2\pi \phi_{ij}(t) [\int_{M} \frac{\partial p(x') }{\partial x}\delta(x-x')\frac{dx'} {(2\pi)^{\frac{d+1}{2}}}+f_i]
 \\ = -2\pi \phi_{ij}(t) [f_i+\frac{\partial p_i(x) }{\partial x}]
= [u_0]_M \gamma^{-1/2} \ g^{-1/2}[x_i(t),x_j(t)]
\end{multline}

Interpreting the result, we find that the velocity field is proportional to the interaction with a force acting on the system, plus the rate of change of pressure with respect to the spatial coordinates. The term \( \phi_{ij}(t) \) acts as a geometric correction for the manifold containing the fluid. As \( A_{ij} \) approaches zero, the term \( \phi_{ij}(t) \) tends toward one, which does not introduce any inconsistencies in the solution derived.

Additionally, the constant \( -2\pi \) indicates that there is some cyclical behavior in the solution, which arises as a remainder from the Fourier Transform applied to the manifold. For an observer situated within the manifold, this framework remains consistent, described by a geometric term given by \( \gamma^{-1/2} \, g^{-1/2}[x_i(t),x_j(t)] \).

The \( D \)-dimensional volume of the manifold, \( V_D \), is associated with a \((D-1)\)-dimensional boundary of the volume or surface , denoted as \( V_{D-1} \), which represents the a $D-1$ volume like a surface of the manifold. The pressure may have written as $p=f \ V_{D-1}^{-1}$. Developing its derivative and replacing it in the equation for the velocity field it gives changing indexes.

\begin{multline}
    u_i[x_i(t)]=-2\pi \phi_{ij}(t)[f_j(1+V_{D-1}^{-2}\frac{\partial V_{D-1} }{\partial x})+(\frac{\partial f_j}{\partial x})V_{D-1}^{-1}]
    \\=-2\pi\exp{[\gamma(2\pi)^{d+1}A_{ij}(t)t}]
     [f_j(1+V_{D-1}^{-2}\frac{\partial V_{D-1} }{\partial x})\\+(\frac{\partial f_j}{\partial x})V_{D-1}^{-1}] =  [u_0]_M \gamma^{-1/2} \ g^{-1/2}[x_i(t),x_j(t)]
\end{multline}
The dimensions can be constructed as the sum of spatial and temporal components, where
\begin{equation}
    D = d_x + d_t = d + 1.
\end{equation}
Here, \(d_x\) represents the spatial dimensions, and \(d_t\) denotes the time dimensions, giving a total dimensionality of \(D\). Analyzing the right side of the equation for the velocity field and writing it as
\begin{equation}
    \frac{d x_i (t)}{dt}= [u_0]_M \gamma^{-1/2} g^{-1/2}[x_j(t),x_i(t)]
\end{equation}
And multiplying for the right by $g^{1/2}[x_j(t),x_i(t)]$.
\begin{equation}
    \frac{d x_i (t)}{dt}g^{1/2}[x_j(t),x_i(t)]= [u_0]_M \gamma^{-1/2}
\end{equation}
This equation must describe the fluid in relation to its own geometry and the geometry of the space-time in which it is contained. If $g[x_i(t),x_j(t)]=g$ is constant and does not depend on the coordinates and the parameter, it becomes possible to integrate the equation and obtain a trivial and known solution.
\begin{equation}
    x_i(t)= v_{i0}t+x_{i0}
\end{equation}
With $v_{i0}=[u_0]_M \gamma^{-1/2}g^{-1/2}$.
Utilizing the inner product for the continuous case as shown in [1] changing the upper limit of the product operator for any $S \in \mathbb{R}$ which is an equivalent, efficient and better way of defining the element of manifold compared to the previously given.

\begin{widetext}
\begin{equation}
    s[^{|\mu|}_{|\alpha|}]=[\langle x[^{|\mu|}] |  x[^{|\alpha|}] \rangle ]^{1/p(D)}=\int_{\tau[^{|\mu|}]}^{\tau'[^{|\mu|}]} \prod_{p(D)\geqslant1}^{p(D)\leqslant S} \ \bigg[g[^{|\mu|}_{|\alpha|}]_{[p(D)-1]} \ \frac{d x[_{|\mu|}]_{[p(D)-1]}}{d\tau[^{|\mu|}]} \ \frac{d x[^{|\alpha|}]_{[p(D)-1]}}{d\tau[^{|\mu|}]} \bigg]^{1/p(D)} d\tau[^{|\mu|}]
\end{equation}
\end{widetext}

This length can possibly never be infinite without loops because of the initial hypothesis of the bounded manifold and it is not necessary it be path commutative.

Even though this equation refers to macrotensors, it is essential to understand that each index of the macrotensor transforms through tensor multiplication by another tensor (a microtensor). A macrotensor can be viewed as a specific type of multilinear application, acting between collections of ordered vector spaces. Each vector space within these collections is, in turn, a set of vector spaces.

The macrotensor, denoted by \( g[^{|\mu|}_{|\alpha|}] \), represents a mapping from one collection of vector spaces to another:

\[
g[^{|\mu|}_{|\alpha|}] : V_0 \otimes V_1 \otimes \dots \otimes V_{d_1} \rightarrow W_0 \otimes W_1 \otimes \dots \otimes W_{d_2}
\]

Here, \( |\mu| = \mu_0 ,\ \mu_1 ,\ \dots ,\ \mu_{d_1} \) represents the combined indices of the output vector spaces \( W_0, W_1, \dots, W_{d_2} \). \( |\alpha| = \alpha_0 ,\ \alpha_1 ,\ \dots ,\ \alpha_{d_2} \) represents the combined indices of the input vector spaces \( V_0, V_1, \dots, V_{d_1} \).

In this context, the entire multilinear application (or macrotensor) can be understood as isomorphic to a ring \((K^D, \otimes, \oplus)\), where \(K^D\) represents the space of all possible tensors with dimension \(D=d_1 \otimes d_2\). The operations \(\otimes\) (tensor product) and \(\oplus\) (direct sum) define the algebraic structure of this space. 

The manifold itself contains elements that can be reduced to specific cases of tensors and applied to this formulation.

\begin{widetext}
\begin{equation}
    s(t',t)=[\langle x_{\mu} |  x_{\alpha}] \rangle ]^{1/p(D)}=\int_{t}^{t'} \prod_{p(D)\geqslant1}^{p(D)\leqslant S} \ \bigg[g^{\mu\alpha}_{[p(D)-1]} \ \frac{d x_{\mu[p(D)-1]}}{d t} \ \frac{d x_{\alpha {[p(D)-1]}}}{dt}\bigg]^{1/p(D)} dt
\end{equation}
\end{widetext}

The element of the manifold defines a trajectory between two coordinates on the manifold at times \( t \) and \( t' \). The integrand can evaluate the path dictated by the velocity field \( u[x(t)] \), where its components must be derived and substituted into the integral. Alternatively, the path can be assessed using the displacement field \( x(t) \), utilizing the established expressions for \( u[x(t)] \).

Given that the manifold is bounded, the length of a path cannot be infinite without forming loops. Consequently, the surface $\Lambda \subseteq M$ under the curve of the element of the manifold is represented by the following integral expression:

\begin{multline}
\int_{\Lambda} s(t',t) , d^Dx \leqslant \int_M \sum_{i=0}^{n} \prod_{j=0}^{m} s_i(t'_1,t_1) s_j(t'_2,t_2) , d^Dx \\ \ 
\leqslant P \in M_{K^D}
\end{multline}

While this integral may yield a large value, it remains constrained within the bounded nature of the manifold. Therefore, even though the manifold is bounded and its limits may vary, it can be considered as a collection of multiple trajectories with loops or a infinite sum and superposition of theses ones that together form the manifold.

The macrotensor structure limits the growth of the integral, and the finite domain ensures that the integral converges to some quantity $P$.

Ultimately, because the length is confined by the boundaries of the manifold, the total integral is also constrained, converging to a specific quantity. Thus, even as the integral is extended from minus infinity to positive infinity across the entire manifold, it remains finite. This confirms that the combination of the macrotensor structure and the boundaries of the manifold guarantees convergence, where this infinity is greater than that for the surface $\Lambda$, yet both are smooth and convergent.

\section{\label{sec:level3} Transformations}

It was seen in the previous section that one possible solution for the Navier-Stokes Equations behaviors as a inertial frame of reference with constant velocity $v_{0i}=[u_0]_M\gamma^{-1/2}g^{-1/2}$ on the manifold M. Where the Poincaré Group and all transformations for this kind of physical systems may be used as usual. Nonetheless, remains questions about transformations between manifolds and how does that affect the observed quantities for the velocity field?

Using the hypothesis that there exists another topological space with a given metric $(G,g')$ and a class of maps $\Sigma: (M,g) \rightarrow (G,g')$ connecting these two manifolds continuously, as denoted in [4], it is assumed the following. For every open set $\omega$ in $(G,g')$, the preimage $\Sigma^{-1}(\omega)$ is open in $(M,g)$. Furthermore, if the inverse map $\Sigma^{-1}: (G,g') \rightarrow (M,g)$ is also continuous, then the bijection $\Sigma$ is a homeomorphism, preserving the topological structure of both manifolds. 

If, in this class of maps, there also exists an isometry $\Omega: (M, g) \rightarrow (G,g')$ that preserves distances via the metrics, as used in [1], such that $s[^{|\mu|}_{|\alpha|}] = s[^{\Omega[|\mu'|]}_{\Omega[|\alpha'|]}]$, and the inverse $\Omega^{-1}: (G,g') \rightarrow (M, g)$ is also differentiable $k$ times ($C^k$) and therefore continuous, then $\Omega$ is a $C^k$-diffeomorphism, preserving the differentiable structures and distances between the manifolds.

Physically, this can be interpreted as a mapping that preserves the geometric and topological properties of the space while allowing the transformation between different metrics onto spaces, allowing symmetries and therefore, conserved quantities, via symmetries groups and other descriptions. But if some of the morphisms here described are not accomplished, being possibly just necessarily the diffeomorphisms for having differentiable structures without matter the changes of geometric and topological structures beetween manifolds also for definition of the velocity vector field as derivative of a displacement field over some parameter of time. So, the constant solution found for $u[x(t)]$ will be just constant for the observers of a given manifold, not preserving its values or not even being constant for the observers of other manifold if the class of maps $\Sigma-\{C^k\}$ are not fulfilling.

One example of this is the next transformation which is a diffeomorphism. Let be $\eta:  (M, g) \rightarrow (G,g')$ in such a way $[u_0]_G$ is constructed from a factor of correction from $[u_0]_M$ given by the argument of $\eta$. So, $ [u_0]_M \thicksim [u_0]_G$ implying,

\begin{multline}
[u_0]_G  = \eta [u_0]_M = (1+\alpha)[u_0]_M = (1+[u_0]^{-1}_M\beta)[u_0]_M \\= [u_0]_M+[u_0]^{-1}_M\beta[u_0]_M =  [u_0]_M + D_{\beta} \ , \ \forall \beta \in K^{d+1}   
\end{multline}

If the expression is

\begin{multline}
    D_{\beta} = [u_0]^{-1}_M\beta[u_0]_M =\beta [u_0]^{-1}_M[u_0]_M\\=[u_0]^{-1}_M[u_0]_M\beta =\beta    
\end{multline}

It gives the next shape to the argument of the vector field.

\begin{equation}
    [u[x(t)]]_G = [[u_0]_M+\beta]\gamma^{-1/2}g^{-1/2}[x_i(t),x_j(t)]
\end{equation}

Being just a rescaling of $[u_0]_M$ for the observers in $(G,g')$ depending on whether or not $\beta$ depends on the coordinates being a local or a global symmetry, respectively. Nonetheless, if $ D_{\beta} = [u_0]^{-1}_M\beta[u_0]_M \not =\beta [u_0]^{-1}_M[u_0]_M =\beta$, the transformation $\eta$ represents naturally a diagonalization of $\beta$.

\begin{equation}
    (M,g)_B \xrightarrow{\mathcal{\eta}} (G,g')_{B'}= (M,g')_{B'}
\end{equation}

If and only if the product $[u_0]^{-1}_M \beta[u_0]_M$ is well defined via the dimensions. This indicates that the transformations between manifolds are actually transformations of basis within the same manifold with different metric, transitioning from a given basis to another constructed from the eigenvectors, which can be represented in diagonal form. In consequence, the vector field is given by 

\begin{equation}
    [u(x(t))]_M = [[u_0]_M + D_{\beta}]\gamma^{-1/2}g'^{-1/2}[x'_i(t),x'_j(t)]
\end{equation}

In this case $\beta = \beta[x'_i(t)]$ necessarily depends on coordinates, indicating a local symmetry on the manifold implying a conserved quantity which makes the other point of view for the expression of $u[x(t)] = -2\pi \phi (f-\partial p)$ (abusing of notation) be conserved on the respective coordinates of this manifold.

This result must hold for equation (2). Since the differentiable structure was preserved in the transformation, the equation has the same argument. Replacing the terms in the expression it provides the next result,

\begin{multline}
\bigg[ \frac{\partial [u_0]_M}{\partial x'_i} + \frac{\partial D_\beta}{\partial x'_i}\bigg] \gamma^{-1/2}g'^{-1/2}[x'_i(t),x'_j(t)] \\ - \frac{1}{2} \bigg[[u_0]_M + D_\beta\bigg]\gamma^{-1/2}g'^{-3/2}[x'_i(t),x'_j(t)] \frac{\partial g'[x'_i(t),x'_j(t)]}{\partial x'_i} = 0
\end{multline}

Multiplying by $\gamma^{-1/2} g^{-3/2}$ and vanishing constant terms.

\begin{multline}
\frac{\partial D_\beta}{\partial x'_i}g'[x'_i(t),x'_j(t)] - \frac{1}{2} \bigg[[u_0]_M + D_\beta\bigg] \frac{\partial g'[x'_i(t),x'_j(t)]}{\partial x'_i} = 0
\end{multline}

Doing the same in the left side of the expression found for the velocity field.

\begin{equation}
\frac{\partial \phi_{ij}(t)}{\partial x'_i}\bigg[f_i + \frac{\partial p_i}{\partial x'_i}\bigg] + \phi_{ij}(t) \left[\frac{\partial f_i}{\partial x'_i} + \frac{\partial^2 p_i}{\partial {x'_i}^{2}}\right] = 0 
\end{equation}

Developing the term for the partial derivative of $\phi$.

\begin{multline}
\frac{\partial \phi_{ij}(t)}{\partial x'_i} =  \gamma (2\pi)^{d+1} \left[\frac{\partial A'_{ij}(t)}{\partial x'_i}t + A'_{ij}(t)\frac{\partial t}{\partial x'_i}\right] 
\\ \exp{\left[\gamma (2\pi)^{d+1} A'_{ij}(t) t\right]} \\=  \gamma (2\pi)^{d+1} \left[\frac{\partial A'_{ij}(t)}{\partial x'_i}t + A'_{ij}(t)\frac{\partial t}{\partial x'_i}\right] \phi_{ij}(t)
\end{multline}

Replacing in the equation (33) and multiplying by $[\phi_{ij}(t)]^{-1}$ the equation gives.

\begin{multline}
\gamma (2\pi)^{d+1} \left[\frac{\partial A'_{ij}(t)}{\partial x'_i}t + A'_{ij}(t)\frac{\partial t}{\partial x'_i}\right]    \Bigg[f_i + \frac{\partial p_i}{\partial x'_i}\Bigg] \\+  \left[\frac{\partial f_i}{\partial x'_i} + \frac{\partial^2 p_i}{\partial {x'_i}^{2}}\right] = 0     
\end{multline}

It indicates a local symmetry with respect to the coordinates  $x'_i$ implying the conservation of linear momentum in the system. These equations are derived from the Navier-Stokes Equations.

In general, numerous transformations for various quantities can be computed based on their dependence on specific coordinates, which affects the system's symmetries, conserved quantities, and governing equations. Notably, there are interesting cases where the transformations may lack continuity, fail to be differentiable, or are conformal.

\section{\label{sec:level4} Discussion and Conclusions}

In this paper, we presented a geometric approach to the Navier-Stokes equations, integrating techniques from Differential Geometry, Fourier Analysis, Abstract Algebra, and Group Theory. This methodology introduces a new class of solutions that effectively capture complex fluid behaviors with insightful physical interpretations, thereby advancing our understanding of fluid dynamics. These findings have implications that extend beyond fluid dynamics, potentially informing fields like engineering, climate modeling, and broader applications in physics and mathematics, particularly where geometry is relevant through the function $\phi_{ij}(t)$ through $A_{ij}(t)$. However, the solutions derived here are constrained by the algebraic approach employed, suggesting the potential for a more general solution class, including relativistic extensions framed in terms of the energy-momentum tensor with series expansions for perturbative analysis.

This approach contributes to ongoing efforts to gain a deeper understanding of the Navier-Stokes equations under specific conditions, building on established partial, weak, and alternative solution frameworks. The alignment between the geometric perspective and weak solutions is particularly significant, as it may shed light on key questions of uniqueness, stability, and smoothness. The results obtained reproduce known solutions, suggesting this approach could serve as a bridge between classical and weak solution methods. Furthermore, the methodology reveals intriguing links between macrotensors and manifold components, opening a new path to exploring solutions and its convergence.

The algebraic structure embedded within collections of vector spaces that define the manifold is dynamic and complex enough to incorporate bounded curvature, differentiability, and boundary contours rather than requiring an infinite expanse. Here, the surface of the manifold itself is less relevant than its contours and boundaries, which represent a sum and superposition of multiple or infinite trajectories where geodesics trace paths that contribute to manifold construction.

The study also underscores connections between Fluid Dynamics and advanced fields like Quantum Field Theory and M-theory, encouraging further cross-disciplinary research. Methods and insights drawn from these areas hint at an adaptable framework for fluid dynamics that could extend to these domains.

The transformation section leverages symmetries across manifolds to derive conserved quantities, offering essential physical insights. This approach enables diagonalizations and basis changes that imply geometric and topological shifts in the fluid, as seen from the perspective of observers situated on different manifolds. Consequently, quantities that appear invariant may exhibit variability when observed from different manifolds, remaining constant only locally. This suggests that conservation may be observer-dependent, contingent upon the manifold, and its specific coordinates.

Physically, this suggests that variations in inertial frames across different manifolds are essential for understanding these fluid dynamics solutions. However, exploring the effects of strong gravitational fields, quantum couplings, alternative formulations of the Navier-Stokes equations, as well as their associated solutions, different morphisms, analogous structures, symmetries, and conserved quantities, presents a promising avenue for future research.

To extend the relevance of these solutions and support potential applications, further simulations, experimental analyses, and continued theoretical developments are essential.

\begin{acknowledgments}

Special thanks go to family and friends, including Lic. Gisela Patricia Céspedes Barrantes, Lic. Marco Vinicio Sacasa Soto, Kiara Grisel Sacasa Céspedes, Óscar R. Jiménez Fernández, Lilliam Barrantes Rodríguez, Yalile Soto Alvarado, M.Sc. Irwin J. Céspedes Barrantes, Octavio Gerardo Rojas Quesada, and many others whom I could never finish listing, for their unwavering belief in me.

Also, a heartfelt acknowledgment to Zoe Natalia Bolaños Segura. Her curiosity and thoughtful reflections throughout our conversations helped shape the transformation section and the interpretation, ultimately guiding the direction of this paper, thank you.

The author is deeply grateful to all those who have supported him in every way and through every situation that has arisen over space-time, enabling progress and improvement. Thank you very much for everything. Each one of you deserves great recognition and credit for this work and others, as you are the ones who have truly made it possible.

\end{acknowledgments}

\end{document}